\documentclass[letter,traditabstract]{aa}  

\usepackage{natbib}
\usepackage{dcolumn,lscape,longtable} 
\bibpunct{(}{)}{;}{a}{}{,}
\usepackage{graphicx}
\usepackage{txfonts}

\newcommand{\ngc}{NGC$\,1999$}

\newcommand{\herschel}{\textit{Herschel}}
\newcommand \msun{\hbox{$\hbox{M}_{\odot}$}}
\newcommand{\micron}{$\mu$m}

\begin{document}

\title{Hier ist wahrhaftig ein Loch im Himmel \thanks{\herschel{} is an
    ESA space observatory with science instruments provided by
    European-led Principal Investigator consortia and with important
    participation from NASA} \thanks{This publication includes
    data acquired with the Atacama Pathfinder Experiment (APEX; 
    proposal E-082.F-9807 and E-284.C-5015). APEX is a collaboration
    between the Max-Planck-Institut f\"ur Radioastronomie, the
    European Southern Observatory, and the Onsala Space Observatory.
    This paper includes data gathered with the 6.5 meter Magellan
    Telescopes located at Las Campanas Observatory, Chile.}  }

\subtitle{{\it The NGC\,1999 dark globule is not a globule
}}


   
   \author{T.~ Stanke\inst{1}\and 
     A.~M.~Stutz\inst{2,3}\and
     J.~J.~Tobin\inst{4}\and
     B.~Ali\inst{5}\and
     S.~T.~Megeath\inst{6}\and
     O.~Krause\inst{2}\and
     H.~Linz\inst{2}\and
     L.~Allen\inst{7}\and
     E.~Bergin\inst{4}\and
     N.~Calvet\inst{4}\and
     J.~Di~Francesco\inst{8,9}\and
     W.~J.~Fischer\inst{6}\and
     E.~Furlan\inst{10}\and
     L.~Hartmann\inst{4}\and
     T.~Henning\inst{2}\and
     P.~Manoj\inst{11}\and
     S.~Maret\inst{12}\and
     J.~Muzerolle\inst{13}\and
     P.~C.~Myers\inst{14}\and
     D.~Neufeld\inst{15}\and
     M.~Osorio\inst{16}\and
     K.~Pontoppidan\inst{17}\and
     C.~A.~Poteet\inst{6}\and
     D.~M.~Watson\inst{11}\and
     T.~Wilson\inst{1}
     }

   \institute{ESO, Karl-Schwarzschild-Strasse 2, 85748 Garching bei M\"unchen, Germany; \email{tstanke@eso.org}
     \and
     Max-Planck-Institut f\"ur Astronomie, K\"onigstuhl 17, D-69117 Heidelberg, Germany
     \and
     Department of Astronomy and Steward Observatory, University of Arizona, 933 North 
     Cherry Avenue, Tucson, AZ 85721, USA 
     \and
     Department of Astronomy, University of Michigan, Ann Arbor, MI 48109, USA
     \and
     NASA \herschel{} Science Center, California Institute of Technology, 770 South Wilson Ave, 
     Pasadena, CA91125, USA
     \and
     Department of Physics and Astronomy, University of Toledo, 2801 West Bancroft Street, Toledo, OH 43606, USA
     \and
     National Optical Astronomy Observatory, 950 N. Cherry Ave., Tucson, AZ 85719, USA 
     \and
     Department of Physics and Astronomy, University of Victoria, P.O. Box 355, STN CSC, Victoria BC, V8W 3P6, Canada
     \and
     National Research Council Canada, Herzberg Institute of Astrophysics, 5071 West Saanich Road, Victoria BC, V9E 2E7, Canada 
     \and
     JPL, California Institute of Technology, Mail Stop 264–767, 4800 Oak Grove Drive, Pasadena, CA 91109, USA
     \and
     Department of Physics and Astronomy, University of Rochester, Rochester, NY 14627, USA
     \and
     Laboratoire d'Astrophysique de Grenoble, Universit\'e Joseph Fourier, CNRS, UMR 571, BP 53, F-38041 Grenoble, France
     \and
     Space Telescope Science Institute, 3700 San Martin Dr., Baltimore, MD 21218, USA
     \and
     Harvard-Smithsonian Center for Astrophysics, 60 Garden Street, Cambridge, MA 02138, USA
     \and
     Department of Physics and Astronomy, Johns Hopkins University, 3400 North Charles Street, Baltimore, MD 21218, USA 
     \and
     Instituto de Astrofisica de Andalucia, CSIC, Camino Bajo de Huetor 50, E-18008, Granada, Spain 
     \and
     Division of Geological and Planetary Sciences 150-21, California Institute of Technology, Pasadena, CA 91125, USA 
   }

   \date{}

 
  \abstract {The \object{NGC\,1999} reflection nebula features a dark patch
    with a size of $\sim$10,000\,AU, which has been interpreted as a small,
    dense foreground globule and possible site of imminent star
    formation.  We present \herschel{} PACS far-infrared 70 and
    160\,\micron{} maps, which reveal a flux deficit at the location
    of the globule.  We estimate the globule mass needed to produce
    such an absorption feature to be a few tenths to a few \msun. Inspired by
    this \herschel{} observation, we obtained APEX LABOCA and SABOCA
    submillimeter continuum maps, and Magellan PANIC near--infrared images
    of the region.  We do not detect a submillimer source at the location
    of the \herschel\ flux decrement; furthermore our observations place an
    upper limit on the mass of the globule 
    of $\sim$2.4$\cdot 10^{-2}$\,\msun.
    Indeed, the submillimeter maps appear to show a flux depression as well.
    Furthermore, the near--infrared images detect faint background stars
    that
    are less affected by extinction inside the dark patch than in its
    surroundings. We suggest that the dark patch is in fact a hole or cavity
    in the material producing the \ngc{} reflection nebula, excavated by
    protostellar jets from the V\,380\,Ori multiple system.
}

   
\maketitle

\section{Introduction}

In the year 1774 C.E.\ Sir Friedrich Wilhelm Herschel first noticed
patches of sky in the constellation Scorpio that were devoid of stars.
Unable to find even the faintest star in these regions, his sister
Caroline reported him to exclaim: ``Hier ist wahrhaftig ein Loch im
Himmel!'' (``Truly there is a hole in the sky here!'').  These dark areas are
now known to be due to obscuring material \citep[e.g.,][]{barnard27}: clouds
of molecular gas, whose dust content absorbs the light of background
stars.  Dark clouds, ranging in mass and size from giant molecular
clouds (GMCs) to tiny globules, are the sites of star formation in our
Galaxy.

\begin{figure}
\begin{center}
  \scalebox{0.325}{{\includegraphics[angle=270]{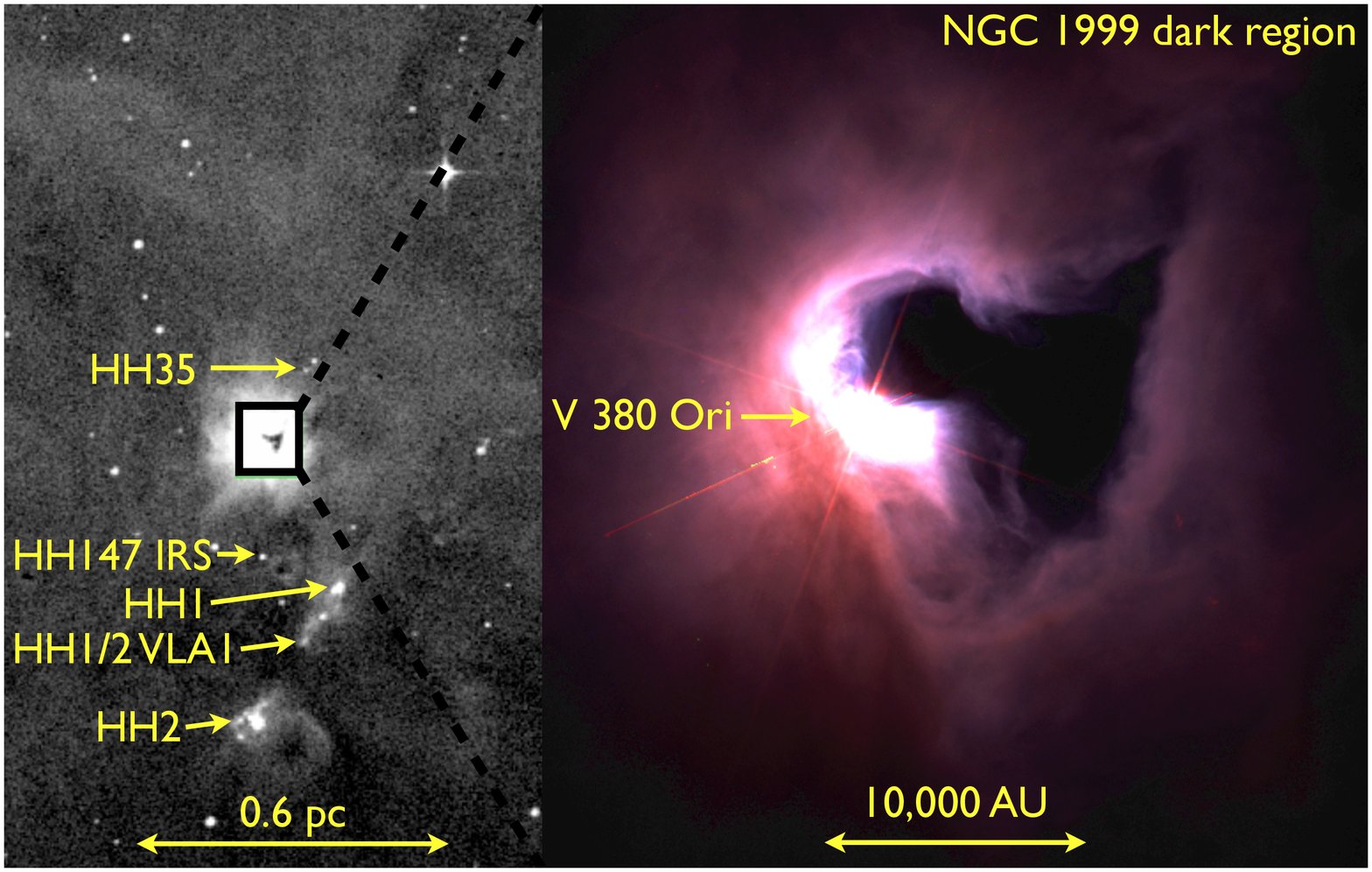}}}
  \caption{\ngc{} and HH\,1/2 region (DSS, left) and HST 
    F450W/F555W/F675W true color image of the \ngc{} dark patch.
    \label{fig:ngc1999_hst}}
\end{center}
\end{figure}

\ngc{} is a small reflection nebula in the \object{LDN\,1641} portion
of the Orion\,A GMC, located in a small group of 22 pre--main sequence
stars and protostars, including the driving source of the prototypical
Herbig--Haro objects \object{HH\,1} and \object{HH\,2} (Megeath et
al., in prep.).  It is illuminated by the Herbig\,Ae/Be star
\object{V\,380\,Ori}, a multiple system with a circumsystem disk where
the primary is a 100\,L$_{\odot}$ B9 star exhibiting strong emission
lines
\citep[e.g.,][]{leinertetal1997,hillenbrandetal1992,alecianetal2009}.
\ngc{} features a compact (20--30\arcsec{}) dark patch (see
Fig.~\ref{fig:ngc1999_hst}), which is described in the literature as a
dark globule and potential site of star formation
\citep[e.g.,][]{herbig1946,herbig1960,warrensmithetal1980}, but which
has never been studied in greater detail before.

We report here the serendipitous and surprising detection with PACS of
a 70 and 160\,\micron{} flux decrement at the location of \ngc{}. 
The maps show a dark patch against the nebulous far--IR emission that
closely resembles the dark patch seen at visible wavelengths.  The MSX, ISO,
and Spitzer space telescopes have detected clouds in absorption at mid--IR
(5-30\,\micron{}) wavelengths against the diffuse IR background of
the Galaxy \citep[e.g.,][]{bacmann00,stutz08,tobin10,stutz09a}, in a few
cases even at 70\,\micron{} \citep{stutz09b}.  However, the detection
of the \ngc{} globule by PACS would be the first detection of a dark
globule in absorption at 160\,\micron{}. At wavelengths $\ge
160$\,\micron{}, cold globules are normally detected through the
emission from cold (10--30\,K) dust.  Furthermore, absorption at these
wavelengths requires extinctions in excess of 100\,$A_V$.

Motivated by the possible discovery of a 160\,\micron{} dark cloud, we
obtained follow--up ground based submillimeter and near--IR
(extinction) observations towards the globule.  We did not detect the
column density or mass of cold dust necessary to produce such an
absorption feature in the far--IR.  We therefore conclude that the
dark globule is actually a cavity in the \ngc{} cloud carved by
outflows from the V\,380\,Ori system.  Thus, the \herschel{}
telescope has discovered what truly is a hole in the sky.

\section{Observations and processing}

\subsection{{\textit{Herschel} PACS observations}}

\begin{figure}
\begin{center}
  \scalebox{0.34}{{\includegraphics[angle=270]{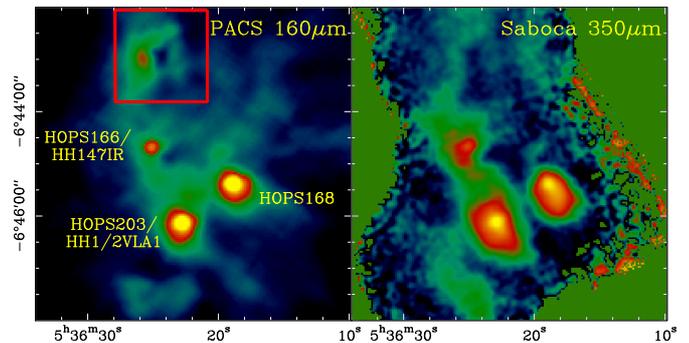}}}
  \caption{NGC\,1999/HH\,1/2 region as seen with
    \herschel{} PACS at 160\,\micron{} (left, the box marks the region shown
    in Fig.\,\ref{fig:img}), and APEX SABOCA at
    350\,\micron{} (right, smoothed to 10\arcsec{} resolution),
    all displayed with a logarithmic color stretch.
    \label{fig:img_160_350_870}}
\end{center}
\end{figure}

\ngc{} was observed with \herschel{} \citep{pilbrattetal2010} with the
PACS instrument \citep{poglitschetal2010} at 70 and
160\,\micron{} on 2009 October 9 (OBSIDs 1342185551 and 1342185552) as
part of the Science Demonstration Phase observations of the
\herschel{} Orion Protostar Survey (HOPS; P.I.\ S.\,T.\,Megeath).  The
data are presented in Figs.\,\ref{fig:img_160_350_870} and
\ref{fig:img} \citep[see also][]{fischeretal2010}.  
They were processed with the \herschel{} Common
Software System (HCSS\footnote{
   HCSS is a joint development by the \herschel{} Science Ground Segment
   Consortium, consisting of ESA, the NASA \herschel{} Science Center, and
   the HIFI, PACS, and SPIRE consortia
})
version 3.0 build 919, following standard steps, using v.\,4 of the calibration
file. After initial processing the two orthogonal scans were combined into a
single observation.  We used the same HCSS software modules as the automatic
PACS pipeline. We used two different approaches for combining the
data into final mosaics:\\
\noindent{\em Method\,1: Simple ("naive") mapping with baseline removal.}
To preserve the extended emission we estimated the sky and the correlated
signal drifts observed in PACS readouts \citep[][]{sauvageetal2010}. To
determine the baselines, we removed pixel--to--pixel electronic offsets 
using the median of the entire time stream, then calculated the median for
each frame/readout. The median values were divided into 500 readout bins, and we
took the minimum value for each bin.  This approach preserves the extended
emission at all spatial scales, but does not remove 1/f noise.
The final mosaic was created using the "photProject" routine.\\
\noindent{\em Method\,2: Optimal mapping with MADmap}\footnote{
   Microwave Anisotropy Dataset mapper \citep[][see \citet{poglitschetal2010}
   for its implementation in HCSS]{cantalupoetal2010}
   } 
to remove the bolometer 1/f noise from the pixel time--line.  The initial
processing is the same as for Method\,1.  However, instead of 'photProject',
MADmap is used to determine an optimal solution for each sky pixel.
This method preserves extended emission {\em and} removes the 1/f noise,
at the price of a reduced sensitivity.

\subsection{Other data}

\noindent {\it APEX SABOCA and LABOCA ---} We obtained submillimeter
continuum maps using LABOCA and SABOCA on APEX.  LABOCA
\citep{siringoetal09} is a $\sim$250 bolometer array operating at
870\,\micron, with a spatial resolution of 19\arcsec.
Observations were done on 2009 November 29 and 30 in fair
conditions, with precipitable water vapour (PWV) values around 1\,mm
($\tau_{{\rm zenith,} 870} \sim 0.26$). We used a combination of
spiral and straight on--the--fly scans in order to recover extended
emission. Data reduction was done with the BOA software (Schuller et
al., in prep.)  following standard procedures, including iterative
source modeling.  SABOCA \citep{siringoetal10} is a 37 bolometer array
operating at 350\,\micron{}, with a resolution of 7\farcs8.
Data were taken on 2009 December 1st and 3rd (\ngc{}) and
on 2008 October 7 and 8 (HH\,1/2 area) with
PWV$<$0.5\,mm ($\tau_{{\rm zenith,} 350} < 1$).  The
observing and data reduction procedures were similar to those used for
LABOCA.  The 350\,\micron{} map is presented in
Fig.\,\ref{fig:img_160_350_870}, and both maps
in Fig.\,\ref{fig:submm} (available in electronic form only).
 
\noindent {\it Magellan PANIC ---} \ngc{} was observed with the PANIC
near--IR camera (2$^\prime$ field of view) at Magellan on 2009 December 4
in photometric conditions through $H$, $K_\mathrm{s}$ and H$_2$ filters
in 0\farcs3 to 0\farcs4 seeing. Separate sky exposures were taken for each
filter.  The data were reduced with standard near--IR routines in the
IRAF UPSQIID package. The photometry is
presented in Appendix \ref{app:panic}.
The $K_\mathrm{s}$ and H$_2$ images are shown in Figs.\,\ref{fig:img} and
\ref{fig:NGC1999_H2}, respectively.

\section{Data analysis}

\subsection{\herschel{} 70\,\micron\ and 160\,\micron}

\begin{figure*}
\begin{center}
  \scalebox{0.63}{{\includegraphics{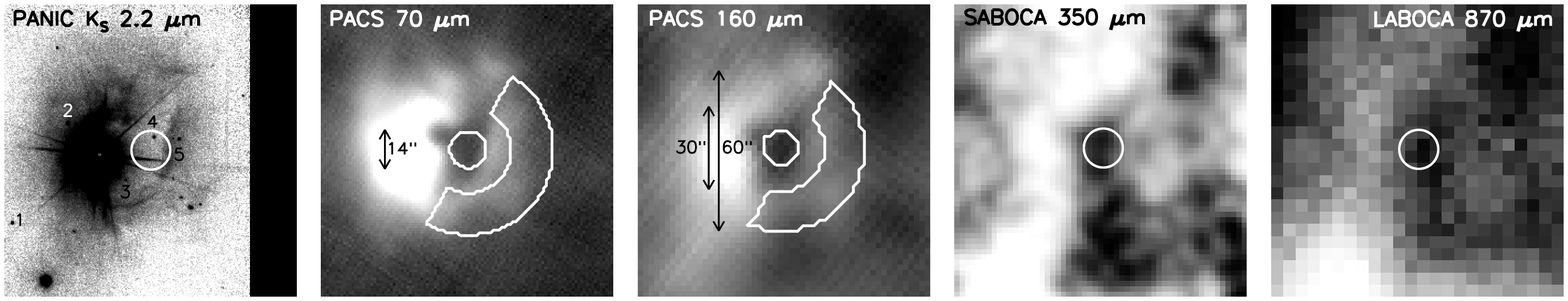}}}
  \caption{1\farcm75$\times$1\farcm75 images of the region
    around \ngc{} (including V\,380\,Ori) centered on R.A.\ = $5^h 36^m
    24.4^s$, Decl.\ = $-06\degr 42\arcmin 56.4\arcsec$ (J2000).  The
    white circle ($14\arcsec$ diameter) on the PANIC near-IR K$_\mathrm{s}$ image
    indicates the \ngc{} region;
    the image is annotated with the source designations for five faint stars
    in the field as listed in Table \ref{table:nir}, where we give their
    ($H - K_\mathrm{s}$) colors.
    The PACS 70\,\micron{} and 160\,\micron{} maps both show
    \ngc{} observed as a flux deficit; the
    overlays represent the pixel masks
    used for the analysis of the \ngc{} feature.  The right-most two
    panels show the SABOCA and LABOCA maps.
    \label{fig:img}}
\end{center}
\end{figure*}

A cursory inspection of the \herschel{} data
(Figs.\,\ref{fig:img_160_350_870} and \ref{fig:img}) immediately reveals
a flux decrement in both PACS maps,
which closely follows the morphology of the putative dark globule in \ngc{}.
Following \citet{stutz09b}, we determine the column density and mass of the
putative globule in the foreground of \ngc{}, assuming that the
observed 70 and 160\,\micron{} decrement is due to extinction.
The optical depth is given by 
\begin{equation}
    \tau = -1.0 \cdot \ln [(f + f_\mathrm{BG})/(f_0 + f_\mathrm{BG})],
  \label{eqn:tau}
\end{equation}
where $f$ is the shadow
flux level, $f_{0}$ is the intrinsic unabsorbed flux level, and
$f_\mathrm{BG}$ is the background contribution to the flux zero--level in the
images.  We estimate the shadow flux level $f$ as the mean
pixel value in a $\sim$7\arcsec{} radius region centered on
the darkest part of \ngc{}; the unobscured flux level $f_{0}$ is
estimated as the median pixel value in a half--annulus with inner and
outer radii of 15\arcsec{} and 30\arcsec{}, with a southwest
orientation chosen to avoid the bright V\,380\,Ori flux.  The $f$
and $f_{0}$ image regions are over--plotted on the PACS maps
in Fig.\,\ref{fig:img}.  The $f_\mathrm{BG}$ flux level is not present in the
\herschel{} maps because constant flux level pedestals
are removed by the data reduction.  We estimate
the 70\,\micron\ BG flux value by interpolating the {\it IRAS}
60 and 100\,\micron\ images; the 160\,\micron\ BG flux values
were obtained from the {\it ISO} Serendipity Survey in the near vicinity of
\ngc{}.  In both cases we take 50\% of the total measured flux as our
estimate for $f_\mathrm{BG}$ as we are only interested in the missing flux
originating from {\it behind} \ngc{}, and not the foreground component.
While this assumption is crude, the
$f_\mathrm{BG}$ values are low compared to the $f$ and $f_{0}$
levels and do not have a large effect on our analysis.

Following Eq.\,\ref{eqn:tau}, we calculate the mean optical depth per pixel. 
Table\,\ref{table:1} presents the results for the data reduced with both methods
(which we find to agree well, indicating that both recover extended
emission equally well). From the resulting optical depth we calculate the
column density and mass (in a $7\arcsec$ radius aperture) required to cause
this flux--decrement at the two PACS wavelengths; these are
presented in Table\,\ref{table:1}.  We use the extreme
\citet{ossenkopf94} model dust opacities for grains with thick ice
mantles: $\kappa_{70} = 541$\,cm$^2$g$^{-1}$ and $\kappa_{160} =
55.7$\,cm$^2$g$^{-1}$, a gas--to--dust mass--ratio equal to 100, and
a distance of 420\,pc \citep[e.g.,][]{mentenetal2007}. 
The resulting column densities and masses for the two wavelength
bands are clearly inconsistent.  For example, we derive that masses of  
0.1\,\msun{} and 2.5\,\msun{} are needed to account for the observed flux  
decrement in the the 70 and 160\,\micron{} data, respectively.

\begin{table}
\begin{minipage}[t]{\columnwidth}
\caption{Far--infrared optical depth, mass, and column density estimates}
\centering
\renewcommand{\footnoterule}{}  
\begin{tabular}{rcccccc}
\hline \hline
$\lambda$ & 
f \footnote{Derived in a $\sim\!7\arcsec$ radius region centered
  on R.A.\ = $5^h 36^m24.4^s$, Decl.\ = $-06\degr 42\arcmin
 56.4\arcsec$, indicated in Fig.~\ref{fig:img} as the inner most
 circular overlay in the \herschel\ PACS images.} & 
f$_{0}$ \footnote{Derived in a $\sim\!15\arcsec$ to $\sim\!30\arcsec$
  radius half--annulus, excluding bright areas associated with V\,380\,Ori, 
  indicated in Fig.~\ref{fig:img} as well.} &
f$_\mathrm{BG}$ \footnote{Estimated background flux contribution to the PACS maps, 
  estimated as 50\% of the measurements from IRAS data 
  (interpolated to 70\,\micron{}) and the 160\,\micron{} {\it
  ISO} Serendipity Survey near \ngc{}.} &
$\tau$ & 
Mass & 
N(H+He) \\
$\mu$m & Jy/$\sq \arcsec$ & Jy/$\sq \arcsec$& Jy/$\sq \arcsec$ &  & \msun & cm$^{-2}$ \\
\hline   
\multicolumn{7}{c}{\it photProject (method 1)}\\
70  & 0.002 & 0.003 & 0.001 & 0.23 & 0.1 & $1.9 \cdot 10^{22}$\\
160 & 0.003 & 0.007 & 0.002 & 0.48 & 2.5 & $3.8 \cdot 10^{23}$\\
\hline
\multicolumn{7}{c}{\it MADmap (method 2)}\\
70  & 0.002 & 0.003 & 0.001 & 0.2  & 0.1 & $1.6 \cdot 10^{22}$ \\
160 & 0.004 & 0.008 & 0.002 & 0.45 & 2.4 & $3.5 \cdot 10^{23}$ \\
\hline                                   
\end{tabular}
\label{table:1}
\end{minipage}
\end{table}

\subsection{Ground-based follow-up}

Neither the APEX 350 nor 870\,\micron{} maps show an emission feature at the
location of the \herschel{} flux decrement.  Instead, the submillimeter
emission bears a strong resemblance to the 160\,\micron{} emission, suggesting
that the 160, 350, and 870\,\micron{} maps are all tracing the  morphology of
the dust emission (Fig.\ \ref{fig:img}). This morphology is suggestive of a
diffuse ring of material, not a hot region absorbed by a foreground cold  
cloud. Assuming optically thin dust emission, total gas$+$dust masses
can be obtained from submillimeter fluxes as
\begin{equation}
M_{\mathrm{G}+\mathrm{D}} = 
\frac{M_\mathrm{G}}{M_\mathrm{D}} \cdot \frac{S_\nu d^2}{B_\nu(T_\mathrm{D}) \kappa_{\nu, \mathrm{D}}},
\end{equation}
where $\frac{M_\mathrm{G}}{M_\mathrm{D}}$ is the gas--to--dust mass ratio, 
$S_\nu$ the flux density, $d$ the distance, $B_\nu(T_\mathrm{D})$ the
Planck function, and $\kappa_{\nu, \mathrm{D}}$ the dust opacity.  

With an rms noise level of $\sim$13\,mJy in our LABOCA map, we should be able
to detect (3\,$\sigma$) a 40\,mJy point source, corresponding to a mass of
0.13\,\msun, assuming $T_{\rm D}=10$\,K and \citet{ossenkopf94}
opacities for thick ice mantle grains. Although the 870\,\micron{}
sensitivity limits are uncertain due to the surrounding highly
structured low surface--brightness emission, we conclude that a mass
around 0.1\,\msun{} should have been detected in the LABOCA map.  More
interestingly, the rms noise on the SABOCA map is around 17\,mJy,
implying a 3$\sigma$ point--source detection limit of $\sim$50\,mJy. Due to the
steeply rising Planck function and dust opacity, this limit corresponds to a
much lower mass detection limit of only $2.4 \cdot 10^{-2}$\,\msun.
For comparison, the mass necessary to produce an extinction shadow as derived
from the 70\,\micron{} map (on the order of 0.1\,\msun) would produce a
350\,\micron{} source with a flux of $\sim$200\,mJy, i.e., a $>10\sigma$
detection. This result is in stark contradiction with the dense globule/shadow
interpretation of the absorption features observed in the \herschel{} data.

\begin{figure}
\begin{center}
  \scalebox{0.49}{{\includegraphics[angle=270.]{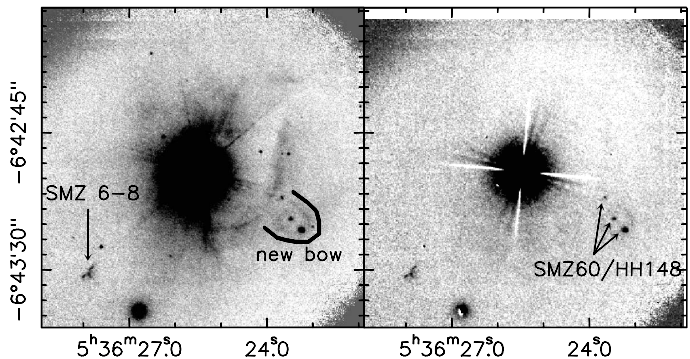}}}
  \caption{Left: PANIC near-IR H$_2$ narrow-band image of the \ngc{};
    right: the same image with a scaled $K_\mathrm{s}$ continuum image 
    subtracted.      
    \label{fig:NGC1999_H2}}                                                         
\end{center}                                                                        
\end{figure}                                                                        

The PANIC images show five faint stars around V\,380\,Ori, marked in
Fig.~\ref{fig:img} with their source designation as listed in Table
\ref{table:nir}, where we also give their $(H-K_\mathrm{s})$ colors and
estimated extinction. Stars 3 and 4 are within the dark patch and have
moderately red colors, implying a mean extinction of $A_V \sim 10$.
In comparison, stars 1, 2, and 5 outside the dark patch are slightly redder
and have a mean extinction of $A_V \sim 12$. None of the five stars are
apparent in the HST images, indicating that they are background stars 
(Fig.\,\ref{fig:ngc1999_hst}). The  $H-K_\mathrm{s}$ colors show that the
extinction toward the dark patch is much less than the 100\,A$_V$ needed to
attenuate the 160\,\micron{} emission and that the extinction through the
dark patch is slightly less than that through the surrounding region.

\section{Discussion}

Optical images of \ngc{} show a dark patch suggestive
of a small, dense globule obscuring the \ngc{} reflection nebula.
Surprisingly, our 70 and 160\,\micron{} images clearly show a  
dark patch against the bright nebula with a strikingly similar  
morphology to that in visible light images.  These \herschel{}  
measurements, combined with subsequent ground based data, lead to the  
conclusion that the dark feature is a hole in the \ngc{} nebula.     
First, we find that the masses needed to cause the 70 and 160\,\micron{}
dark patch --- 0.1 and 2.5\,\msun{}, respectively --- are inconsistent.
Furthermore, the globule is not detected in emission at 350 and 870\,\micron{}:
the SABOCA obervations place an upper limit of $2.4\cdot 10^{-2}$\,\msun{} for
a temperature of 10\,K.  This upper limit is far below the amount of mass
needed to cause the obscuration in the PACS data.  Finally, near--IR
observations with PANIC detect background stars toward the  dark patch; the
$H-K_\mathrm{s}$ colors of these stars are slightly bluer than stars detected
outside the globule, suggesting a lower extinction toward the dark patch. 
Furthermore, the extinctions of the background stars are less than
that required to absorb the 160\,\micron{} flux. Taken together, these  
observation show that the dark patch is not a globule, but instead a  
cavity in the nebula.

The presence of a well delineated cavity of the size of
$\sim$10,000\,AU deserves some attention. With the typical turbulent
velocities on the order of a few km/s in clouds, such a cavity should
be filled on timescales of at most a few 10,000\,yrs and quickly disappear.
The PANIC H$_2$ narrow band data (Fig.\,\ref{fig:NGC1999_H2}) deliver important
hints about the possible origin and peculiar shape of the cavity. They
reveal a previously unknown, faint H$_2$ bow shock enveloping the
SMZ\,60/HH\,148 compact knots \citep{stankeetal2002,corcoranray1995},
constituting the clearest evidence so
far for a collimated flow running northeast to southwest
through the cavity. This flow, which likely originates in the V\,380\,Ori
multiple system, could possibly excavate the southern part of the cavity.
\object{SMZ\,6-8} in the southeastern corner of the PANIC image
resembles a small bow shock in a flow coming from the northwest;
together with \object{HH\,35}, located northwest of V\,380\,Ori
(Fig.\,\ref{fig:ngc1999_hst}), it indicates a second, northwest to
southeast oriented flow, which could be responsible for digging the
northwestern lobe of the cavity.

We note a similar flux depression in the 160\,\micron{} \herschel{}
image southwest of the protostar HOPS\,166, which drives the
\object{HH\,147} outflow \citep{corcoranray1995}.
Optical images (e.g., Fig.\,\ref{fig:ngc1999_hst}) show a circular
reflection nebula marking the HH\,147 outflow cavity, coinciding with
the far--IR flux depression. The \ngc{} dark patch may therefore only
be a somewhat peculiar example of a cavity carved in the ambient medium by an
outflow, rendered particularly visible by the illumination and heating
of the cavity walls by V\,380\,Ori. Sensitive far--IR maps taken with
\herschel{} may therefore provide a new tool to assess the importance
of outflow feedback on cloud cores. 

\begin{acknowledgements}
We thank Frank Bertoldi and Markus Albrecht for their invaluable help with
BOA, Giorgio Siringo for his help with the SABOCA data, Jonathan Williams
for encouraging discussions, and the APEX staff for their help wiht taking the
data. Based in part on observations made with \herschel{}, a European
Space Agency Cornerstone Mission with significant participation
by NASA. Support for this work was provided by NASA through an award issued by  
JPL/Caltech. JJT acknowledges funding through HST-GO-11548.04-A. 
Fig.\,\ref{fig:ngc1999_hst} produced from data taken
with the NASA/ESA Hubble Space Telescope, and obtained from the Hubble
Legacy Archive, which is a collaboration between the Space Telescope
Science Institute (STScI/NASA), the Space Telescope European
Coordinating Facility (ST-ECF/ESA) and the Canadian Astronomy Data
Centre (CADC/NRC/CSA).
\end{acknowledgements}

\bibliographystyle{aa}
\bibliography{14612}

\Online

\begin{appendix}

\section{SABOCA and LABOCA submillimeter maps}
\label{app:submmmaps}

\begin{figure*}
\begin{center}
  \scalebox{0.71}{{\includegraphics[angle=270.]{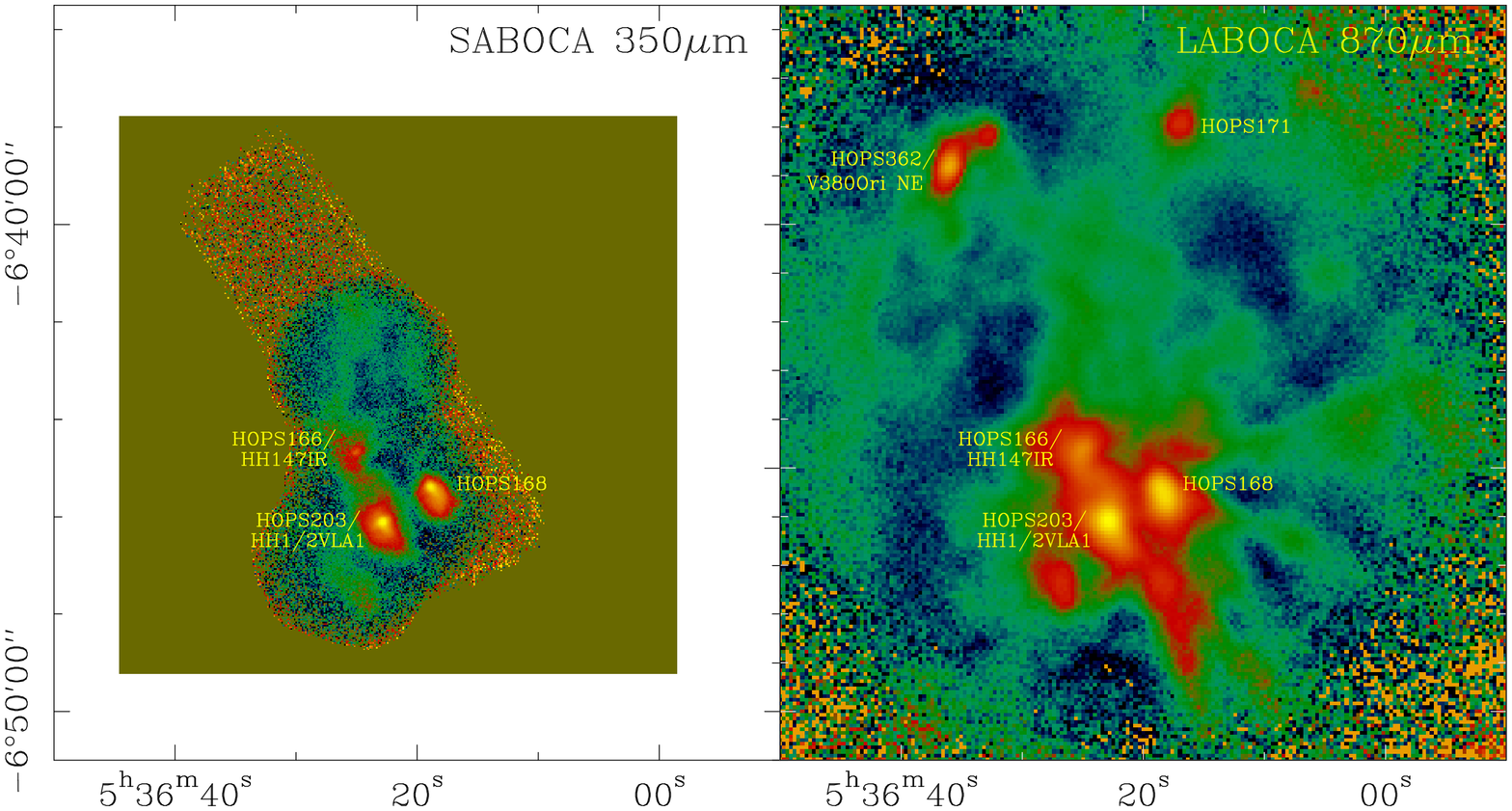}}}
  \caption{Full submillimeter maps. Left: SABOCA, displayed at the original
    resolution of 7\farcs8; right: LABOCA. Protostars in the field are labeled
    with their HOPS designations. For both maps a logarithmic
    color scale is used.
    \label{fig:submm}}                                                         
\end{center}                                                                        
\end{figure*}                                                                        

\section{PANIC $H$ and $K_\mathrm{s}$ photometry}
\label{app:panic}

\begin{table*}
\begin{minipage}[]{\columnwidth}
\caption{PANIC \ngc\ near--IR colors of background stars}
\centering
\renewcommand{\footnoterule}{}  
\begin{tabular}{ccccccc}
\hline \hline
Source \footnote{See Fig.~\ref{fig:img} for source designation.} &
RA (J2000)& 
Decl. (J2000)&
$K_S$ &
$\sigma(K_S)$& 
$(H - K_S)$ &
$\sigma(H - K_S)$ \\ 
  & (h m s) &  ($\degr$ $\arcmin$ $\arcsec$) & mag & mag & mag & mag \\
\hline   

1 & 05 36 27.54 & $-$06 43 22.29 & 16.08 & 0.01 & 0.83 & 0.01\\ 
2 & 05 36 26.19 & $-$06 42 46.23 & 15.68 & 0.01 & 0.95 & 0.02\\ 
3 & 05 36 24.78 & $-$06 43 06.72 & 18.36 & 0.10 & 0.67 & 0.14\\ 
4 & 05 36 24.09 & $-$06 42 51.12 & 16.62 & 0.01 & 0.84 & 0.02\\ 
5 & 05 36 23.48 & $-$06 42 51.86 & 16.71 & 0.02 & 0.95 & 0.02\\ 
\hline                                   
\end{tabular}
\label{table:nir}
\end{minipage}
\end{table*}

Five faint stars are visible in the PANIC $H$ and $K_\mathrm{s}$ images.
Their positions and photometry are listed in Table \ref{table:nir}. 
Photometric calibration was performed with Two Micron All Sky Survey
(2MASS\footnote{
   The Two Micron All Sky Survey is a joint project of the University of
   Massachusetts and the Infrared Processing and Analysis Center/California
   Institute of Technology, funded by the National Aeronautics and Space
   Administration and the National Science Foundation.
   })
photometry of comparison fields containing 20 2MASS sources and verified with
a 2MASS star in the science field. Photometry was conducted with the IRAF task
apphot.

The measured $(H - K_S)$ color was used to estimate the extinction
$A_V$ toward the stars. We assumed an intrinsic $(H - K_S)$ color of 0.17,
derived as the median $(H - K_S)$ colors of stars in an unreddened
control field near Orion (Megeath et al., in prep). We adopted the reddening law
of \cite{indebetouwetal2005} to convert $(H - K_S)$ excess into $K$-band
extinction $A_K$, from which we obtained the optical extinction as 
$A_V = 8 \cdot A_K$, corresponding to a total to selective extinction 
R$_V$ between the diffuse ISM value of 3.1 and the higher values (up to 5)
found for dense clouds \citep[e.g.][]{campeggioetal2007}.

\end{appendix}

\end{document}